\documentclass{article}
\usepackage{arxiv}

\usepackage{graphicx,amsmath,amsfonts,amscd,amsthm,amssymb,pstricks}
\usepackage{color}
\usepackage{dcolumn}
\usepackage{bm}
\usepackage{longtable}
\usepackage{mathrsfs}
\usepackage{textcomp}
\usepackage{esvect}
\usepackage{float}
\usepackage[USenglish,british]{babel}
\usepackage{environ}
\usepackage{xifthen}
\usepackage{xargs}		
\usepackage{hyperref}
\usepackage[separate-uncertainty = true]{siunitx}
\usepackage{physics,mathtools}
\usepackage{multirow}
\usepackage{comment}
\usepackage{enumerate}
\usepackage{appendix}
\usepackage{tikz}	
\usetikzlibrary{arrows, calc, matrix, patterns, decorations.markings, shapes, decorations.pathmorphing, shadows.blur}
\usepackage[utf8]{inputenc}
\usepackage{enumitem}

\usepackage{siunitx}

\title{Measurement of transverse beam emittance of split beams for the CERN Proton Synchrotron Multi-Turn Extraction}

\author{G. Russo \\
CERN, Esplanade des Particules 1, 1121 Meyrin, Switzerland\\
GSI Helmholtzzentrum f\"ur Schwerionenforschung GmbH, Planckstraße 1, 64291 Darmstadt, Germany\\
\And
F. Cerutti \\
CERN, Esplanade des Particules 1, 1121 Meyrin, Switzerland \\
\And
L.S. Esposito \\
CERN, Esplanade des Particules 1, 1121 Meyrin, Switzerland \\ 
G. Franchetti \\
GSI Helmholtzzentrum f\"ur Schwerionenforschung GmbH, Planckstraße 1, 64291 Darmstadt, Germany\\
\And 
M. Giovannozzi\thanks{Corresponding author: massimo.giovannozzi@cern.ch}\\
CERN, Esplanade des Particules 1, 1121 Meyrin, Switzerland \\
\And 
J.R. Hunt \\
CERN, Esplanade des Particules 1, 1121 Meyrin, Switzerland \\
\And
A. Huschauer \\
CERN, Esplanade des Particules 1, 1121 Meyrin, Switzerland}

\begin{document}
\maketitle

\begin{abstract}
Crossing a horizontal nonlinear resonance is the approach that can be used to split a beam in several beamlets with the goal to perform multi-turn extraction from a circular particle accelerator. Such an approach has been successfully implemented in the CERN Proton Synchrotron and is used routinely for the production of high-intensity proton beams for fixed-target physics at the Super Proton Synchrotron. Recently, thanks to the deployment of diamond detectors, originally installed to monitor the beam losses at extraction, it has been possible to measure the horizontal beam emittance of the split beam just prior to extraction. This is the first time that an emittance measurement is attempted for split beams, i.e. in a regime of highly nonlinear beam dynamics. In this paper, the technique is presented and its application to the analysis of the experimental data is presented and discussed in detail. This result is essential for the performance assessment of the splitting process and for the design of further performance improvements.  
\end{abstract}

\keywords{Accelerator modelling and simulations (multi-particle dynamics, single-particle dynamics), Beam dynamics, Instrumentation for particle accelerators and storage rings - high energy (linear accelerators, cyclotrons, electrostatic accelerators)}





\section{Introduction} \label{sec:intro}

At the beginning of 2002, investigations were launched to find a replacement for the so-called continuous transfer (CT)~\cite{Bovet:880676} extraction from the CERN Proton Synchrotron (PS). This extraction mode was proposed in the 1970s and, until 2015, has been the default extraction method for the high-intensity proton beams for the Super Proton Synchrotron (SPS) fixed-target physics programme. The main reason for the replacement of CT was the high level of beam loss at extraction in a significant fraction of the PS ring~\cite{Gilardoni:EPAC08-THPC047,PhysRevSTAB.14.030101}. 

The CT replacement, named Multi-Turn Extraction (MTE)~\cite{PhysRevLett.88.104801,PhysRevSTAB.7.024001}, is based on the generation of a stable fourth-order resonance that is crossed by means of a slow variation of the horizontal tune. This novel manipulation allows particles to be trapped inside the stable resonance islands, so that the initial single-Gaussian beam is split into Gaussian beamlets. Four beamlets are generated by the particles trapped inside the resonance islands, and a fifth one, the core, is made of the beam that remains around the closed orbit after the splitting process. 

This beam manipulation exploits the power of nonlinear dynamics, in particular the effects from sextupole and octupole magnets, which are instrumental in the generation of the stable resonance, and of adiabatic transformation of this nonlinear system that is induced by the horizontal-tune variation to induce the resonance crossing. These two elements enable for virtually lossless beam extraction from the PS. 

The two performance indicators for MTE are the intensity sharing between the core and the four beamlets and the emittance reduction for the core and the beamlets. The initial emittance should be reduced by the splitting process, unless nonlinear effects create an emittance blow-up. Ideally, each beamlet should carry the same beam intensity and horizontal emittance, corresponding to $0.2$ of the initial value when it splits into four beamlets. However, while the intensity sharing can be easily measured by a beam current transformer, the horizontal emittance after splitting cannot be easily ascertained because of the strong nonlinear effects that are at the heart of MTE. The horizontal emittance can of course be measured in the downstream machine, but imposes important constraints during the development and optimisation phases as it required the downstream machine to be available for such studies.

The beam extraction proper~\cite{Giovannozzi:987493} is carried out by means of a set of slow dipoles that generate a closed orbit deformation on a time scale of a few thousand turns in the region around the magnetic extraction septum. Furthermore, fast dipoles, called kickers, with a rise time much shorter than a single turn\footnote{The PS revolution period is \SI{2.1}{\micro s} and the rise time of the kickers is either \SI{56}{ns} or \SI{350}{ns} for the core or the island, respectively.}, further change the closed orbit over five turns for actual beam extraction.  

We recall that the SPS imposes that the split beam is debunched in the PS prior to extraction\footnote{The beam is actually partially re-captured before extraction, using \SI{200}{MHz} cavities.}. As a result, unavoidable beam losses occur during the rise time of the extraction kickers, when the beamlets sweep through the blade of the extraction septum. To mitigate this, a copper blade, called a dummy septum, is installed upstream of the extraction septum to shadow the septum blade~\cite{Bartosik:1463324,Bertone:1697680}. In this way, the extraction losses are localised in the single section of the PS ring where the dummy septum is located, surrounded by appropriate shielding and thus protecting the neighbouring hardware from irradiation. It should be noted that the creation of a barrier bucket has recently been tested~\cite{Vadai:2702852,Vadai:IPAC19-MOPTS106,Vadai:IPAC19-MOPTS107,PhysRevAccelBeams.25.050101} in order to eliminate extraction losses, making MTE a real lossless extraction process. 

After a long period of theoretical and experimental studies~\cite{PhysRevSTAB.9.104001,PhysRevSTAB.12.014001}, including two phases of hardware installation~\cite{Giovannozzi:987493,Bertone:1697680}, by the end of 2015, MTE was fully operational~\cite{Borburgh:2137954,PhysRevAccelBeams.20.061001}. Since then, this complex beam manipulation has been continually developed to improve its performance~\cite{PhysRevAccelBeams.20.014001,PhysRevAccelBeams.20.061001,PhysRevAccelBeams.22.104002,Vadai:2702852} to prepare it for future challenges~\cite{DeLellis2015,Alekhin_2016}. 

In this paper, we report on the first successful measurement of the horizontal emittance for the MTE beam at extraction. In the past, wire scanners~\cite{Hancock:234723,Agoritsas:319333,Steinbach:275924} were used to measure the horizontal profile of the split beam. However, knowledge of the measured profile is not enough to extract information on the value of the beam emittance of the four beamlets and of the core. The four beamlets and the core, which are separated in phase space, might nevertheless be superimposed when projected along the horizontal axis, unless their orientation in the horizontal phase space is particularly favourable at the location of the wire scanner. In general, it is therefore hard to reliably estimate the parameters (mean value and rms) of the five Gaussian profiles using a multi-Gaussian fit. In addition, the strongly nonlinear motion inside the resonance islands implies that knowledge of the projection along the horizontal axis is not enough to estimate the value of the emittance. All these obstacles can be overcome using the temporal signals of beam losses during beam extraction measured by diamond detectors~\cite{ref:pCVD}, as discussed in the remainder of this article. 

The structure of the paper is the following: In Section~\ref{sec:method} a brief overview of the PS ring with the key elements of MTE is provided and, most importantly, the method to extract information about the beam sigma from the time profile of the losses is discussed. In Section~\ref{sec:sim}, the performance of the diamond detector is studied using dedicated numerical simulations that show the linearity of the instrument, an essential feature of the proposed method. In Section~\ref{sec:emitt}, the reconstruction of the beam emittance for the beamlets and the core is presented, and the results are discussed in detail, including the dependence on the non-linear model of the PS ring. Finally, in Section~\ref{sec:perf}, the implications on the MTE performance of these first successful emittance measurements are analysed in detail, with some conclusions drawn in Section~\ref{sec:conc}. 

\section{Emittance measurement technique} \label{sec:method}
\subsection{Overview of the PS ring}
\begin{figure}[!htb]
   \centering
   \includegraphics[trim=25truemm 15truemm 25truemm 5truemm,width=0.7\columnwidth,clip=]{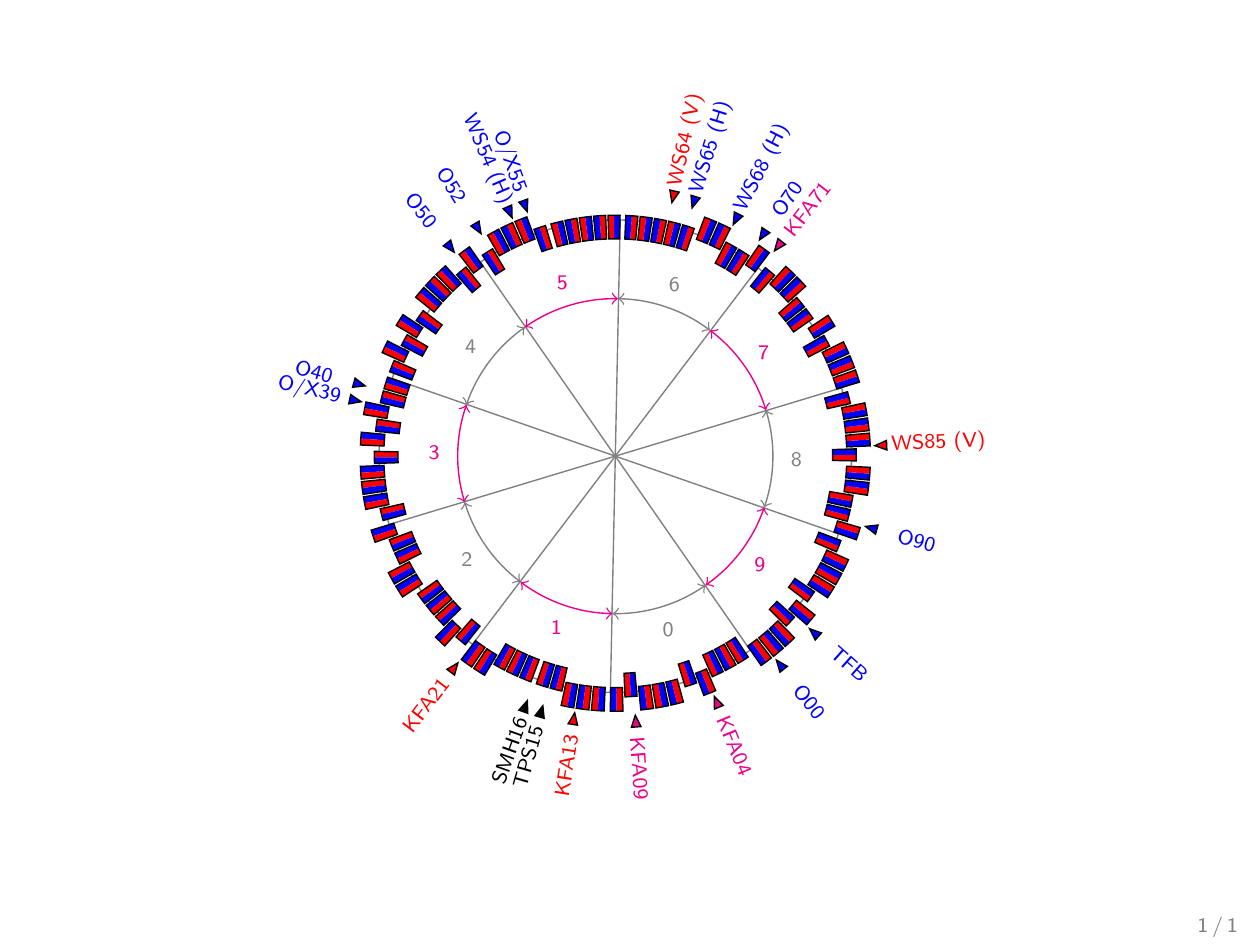}
   \caption{Layout of the PS ring including the combined-function main magnets (red and blue rectangles for the focusing and defocusing parts, respectively) and the location of the key elements for MTE. The ten different sectors of the PS are indicated by the numbers and the radial offset between the rectangles accounts for the four different magnet types (see~\cite{Burnet:1359959} for a recent more detailed information). For the wire scanners (elements with a name starting with WS), the measurement plane is marked in brackets. The number in the element name indicates the straight section number.}
   \label{fig:PS_ring}
\end{figure}

The CERN PS is based on a combined function design~\cite{Regenstreif:214352,Regenstreif:102347,Regenstreif:278715,Regenstreif:342915,Regenstreif:342916}, with a hundred main magnets, each made of two half-units, one focussing and one defocusing, arranged in a regular lattice with superperiodicity ten (see also~\cite{Burnet:1359959}). A sketch of the PS ring is shown in Fig.~\ref{fig:PS_ring}, where the details of the main magnet configuration are clearly visible, as well as the ten sectors.

In the same figure, the key elements for MTE are also shown. Ring elements with names starting with X or O are the sextupoles and octupoles used to create the stable islands. It is important to recall that the main PS magnets are equipped with additional coils, the so-called pole-face-windings (PFW) and figure-of-eight (F8L) loop~\cite{Burnet:1359959}, which have been installed for easier control of the working point. Originally, these additional coils were powered by three independent circuits, allowing control of three of the four parameters $Q_x, Q'_x, Q_y, Q'_y$. In view of the implementation of MTE, an improved powering scheme allowed the control of five currents~\cite{Steerenberg:EPAC06-MOPCH097}. This change was required to ensure the control of the amplitude detuning generated by the octupolar component created by the PFWs. The non-linear components of the main magnets are therefore an essential element of the non-linear dynamics of MTE.

The transverse profiles can be measured by means of wire scanners (elements with a name beginning with WS). A transverse feedback system (TFB)~\cite{Blas:IPAC13-WEPME011,Sterbini:HB14-MOPAB10} provides excitation when the resonance is crossed to increase the probability of a particle being trapped in stable islands~\cite{Sterbini:IPAC16-THPMR043,PhysRevAccelBeams.20.061001}. Kicker dipoles (KFA) are used to create fast deformation of the closed orbit and to extract the beam using the magnetic extraction septum (SMH16). The dummy septum (TPS15) is also shown.  

\begin{figure}[!htb]
   \centering
   \includegraphics[trim=0truemm 2truemm 0truemm 2truemm,width=0.8\textwidth,clip=]{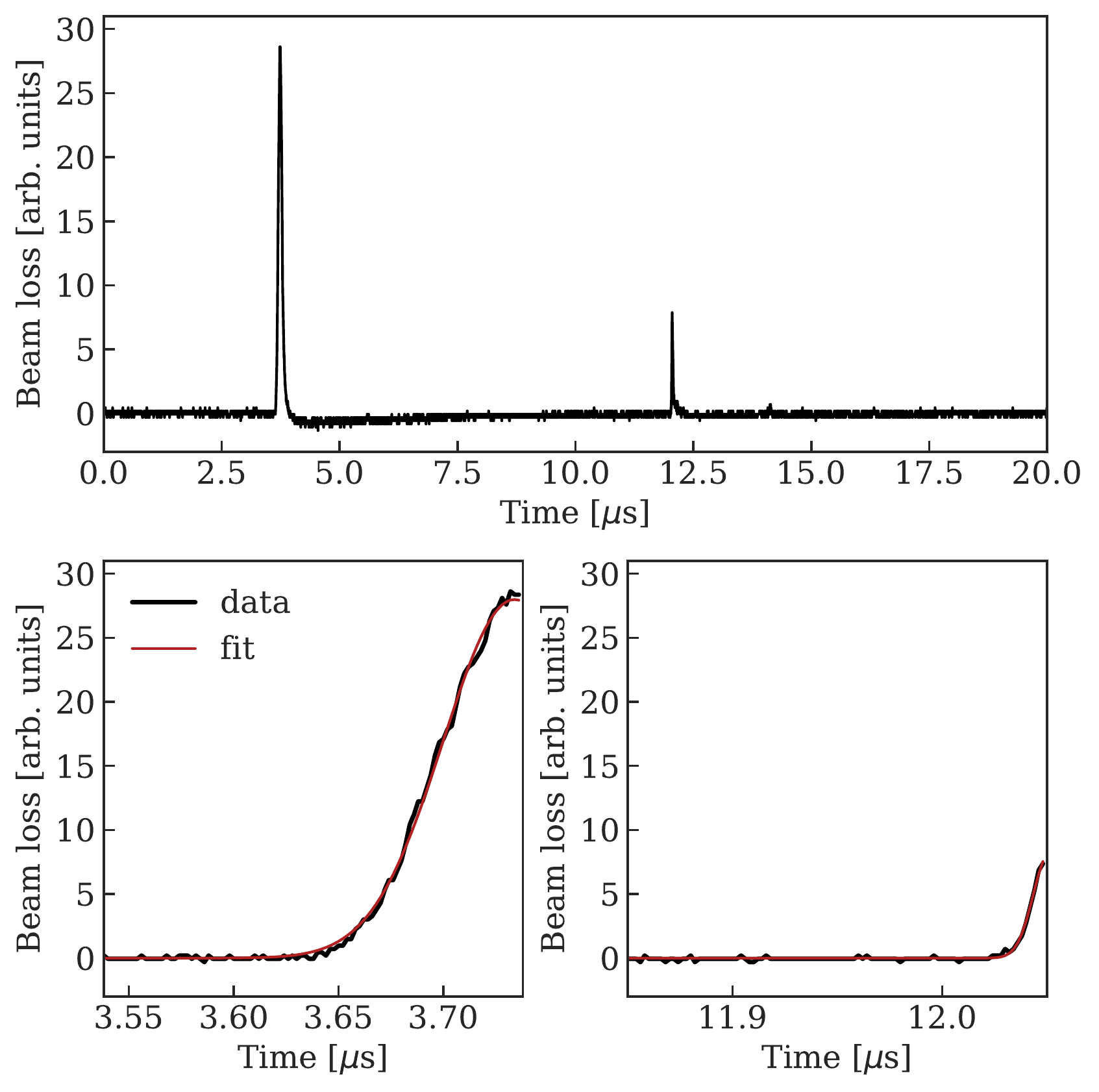}
   \caption{Top: Example of the measured diamond detector signal $\ell_\mathrm{m}(t)$ recorded in SS15. The losses generated by the island sweeping through the blade of the dummy septum cause the first spike, while the second spike represents the losses of the core. Bottom: detailed views of the island and core losses (left and right plot, respectively) together with a fitted Gaussian model (from~\cite{PhysRevAccelBeams.22.104002}).}
   \label{fig:fast_BLMs}
\end{figure}

Two polycrystalline diamond detectors (pCVD)~\cite{ref:pCVD} have been installed on top of the main magnet units just downstream of the straight sections (SS) 15 and 16, where the dummy and extraction septa are installed. These devices are intended to measure the beam losses during the rise time of the extraction kickers. Furthermore, they are capable of providing turn-by-turn losses with a subturn sampling rate, typically at the nanosecond scale, which is a key feature for the proposed emittance measurement method. Due to this characteristic, these detectors allow one to distinguish between the losses generated by the four beamlets and the core when they sweep through the blade of the dummy or extraction septum. An example of the raw signal recorded with the diamond detector in SS15 is shown in Fig.~\ref{fig:fast_BLMs}. 
The difference in peak height is generated by the different rise times of the kickers used to extract the island and the core~\cite{PhysRevAccelBeams.20.061001}, with the time difference between the two peaks corresponding to four PS turns.
\subsection{The proposed method}
In~\cite{PhysRevAccelBeams.22.104002}, a possible method was described to reconstruct the transverse beam distribution, starting from an accurate measurement of losses as a function of time. In this paper, the method is further developed, and the process of beam-matter interaction is simulated in detail to assess the actual impact on the measured data. All this opens up the possibility of reconstructing the horizontal emittance of the core. For the island, additional aspects have to be considered given the highly nonlinear dynamics, and these aspects will be addressed later.

The horizontal beam distribution is described as 
\begin{equation}
\rho(x,t)= \frac{N_{\rm p}}{\sqrt{2\, \pi} \, \sigma} \exp \left [-\frac{1}{2} \left ( \frac{x-\mu(t)}{\sigma} \right )^2 \right ]    \, ,
\end{equation}
where $N_{\rm p}$ is the total intensity of the beam, $\sigma$ the size of the beam and $\mu(t)$ the displacement during the kicker rise time, which can be described as~\cite{Giovannozzi:987493}
\begin{equation}
\mu(t)= \frac{A}{2} \left \{ \sin \left [ 2 \left (t -\frac{\pi}{4} \, \zeta  \right ) \frac{1}{\zeta} \right ] +1 \right \}, \,  0 \leq t \leq \frac{\pi}{2} \zeta \, , 
\label{eq:risetime}
\end{equation}
where $A$ is the maximum horizontal beam displacement at the location where the losses are measured and
\begin{equation}
\zeta = \frac{\tau_\alpha}{\arcsin{(1-2\alpha)}}     \, ,
\end{equation}
where $\tau_\alpha$ is the time during which the kicker sweeps the particles through the amplitude interval $[\alpha \, A , (1-\alpha) \, A]$. It is customary to use $\alpha=0.1$ with $\tau_{0.1}$ is \SI{350}{ns} for the kickers extracting the islands or \SI{56}{ns} for those that extract the core~\cite{Giovannozzi:987493}. 

The number of protons lost $\ell(t)$ due to interaction with the blade of the dummy septum can be obtained under the assumption that a proton hitting the blade is absorbed. In this case, one obtains the following
\begin{equation}
\begin{split}
\ell(t) = \eta \frac{N_{\rm p}}{\sqrt{2 \, \pi} \, \sigma } \,  \frac{\tau}{\tau_{\rm rev}} \int_{x_{\rm s}^-}^{x_{\rm s}^+} \exp \left [-\frac{1}{2} \left ( \frac{x-\mu(t)}{\sigma}\right )^2 \right ] \, \mathrm{d} x \, , 
\end{split}
\label{eq:losses}
\end{equation}
where $\eta$ expresses the fraction of the initial beam intensity that is in the core or in the island, $x_{\rm s}^\pm= x_{\rm s} \pm \Delta/2$ and $x_{\rm s}$ and $\Delta$ are the central position and width, respectively, of the dummy septum blade, $\tau_{\rm rev}$ is the revolution time and $\tau$ is the time interval during which the beam crosses the dummy septum blade. The factor $\tau /\tau_\mathrm{rev}$ corresponds to the integration of the longitudinal distribution, which is assumed to be uniform. Note that $x_{\rm s}= \hat{x}_{\rm s}-\hat{x}_{\rm b}-\hat{x}_{\rm co}$, where $\hat{x}_{\rm s}$, $\hat{x}_{\rm b}$, and $\hat{x}_{\rm co}$ represent the position of the dummy septum with respect to the zero closed orbit, the amplitude of the slow bump, and the closed orbit value, respectively. These quantities differ for the island and the core. 

Equation~\eqref{eq:losses} can be simplified by introducing two approximations. The first one consists of replacing the integral by the function evaluated in the middle of the integration interval, which is a well-justified assumption given the width of the dummy septum blade. The second is based on the observation that losses occur when the function $\mu(t)$ of Eq.~\eqref{eq:risetime}, which expresses the amplitude of the extracted beam as a function of time, is linear. This makes it possible to replace the sinus by its argument, thus obtaining
\begin{equation}
\ell(\hat{t}) \approx \eta \frac{N_{\rm p} \Delta}{\sqrt{2 \, \pi} \, \sigma } \,  \frac{\tau}{\tau_{\rm rev}} \exp \left [-\frac{1}{2} \frac{A^2}{\zeta^2} \frac{\hat{t}^2}{\sigma^2} \right ] \, ,
\label{eq:losses1}
\end{equation}
where the origin of time has been shifted according to $t=\hat{t} + t^\ast$ and $t^\ast$ satisfies the relation $x_\mathrm{s}-A\left ( t^\ast - \pi/4 \, \zeta \right )/\zeta=0$.

This shows that the measured beam losses are represented by a Gaussian function. Hence, the fitted Gaussian features an rms value $\sigma_t$ that satisfies the following relation 
\begin{equation}
    \sigma = \frac{A}{\zeta} \sigma_t \, .
    \label{eq:sigma}
\end{equation}

Therefore, $\sigma$ can be evaluated from Eq.~\eqref{eq:sigma} from knowledge of $\ell_\mathrm{m}(t)$ (obtained from beam measurements), $\mu(t)$ (obtained from the characteristics of the kicker magnets and of the ring optics) and the application of the model~\eqref{eq:losses1}. Knowing $\sigma$ allows the determination of the emittance of the core and island by assuming knowledge of the optical functions of the core and island and the dispersion function. It should be noted that the underlying assumption is that the diamond detector has a linear response to the number of protons hitting the blade of the dummy septum, so $\ell_\mathrm{m}(\hat{t})$ is only a linear transformation of $\ell(\hat{t})$, without non-linear distortions that would affect the determination of $\sigma_t$. 

In this framework, it is possible to introduce two additional quantities, namely the total losses $\ell_\mathrm{tot}$ and the peak of the losses $\ell_\mathrm{peak}$, which are defined as 
\begin{equation}
    \begin{split}
    \ell_\mathrm{tot} & = \int_0^{\frac{\pi}{2}\zeta} \ell (t) \mathrm{d} t \approx \eta \frac{N_{\rm p} \Delta}{\sqrt{2 \, \pi} \, \sigma } \,  \frac{\tau}{\tau_{\rm rev}} \int_{-t^\ast}^{-t^\ast + \frac{\pi}{2} \zeta} \exp \left [-\frac{1}{2} \frac{A^2}{\zeta^2} \frac{\hat{t}^2}{\sigma^2} \right ] \mathrm{d} \hat{t} \, , \\
    & \\
    \ell_\mathrm{peak} & = \ell(0) = \eta \frac{N_{\rm p} \Delta}{\sqrt{2 \, \pi} \, \sigma } \,  \frac{\tau}{\tau_{\rm rev}} \, .
    \end{split}
\end{equation}
It is easily seen that both quantities are linear in beam intensity $N_\mathrm{p}$ and that their ratio depends only on $A, \zeta, \sigma$, and that this ratio differs for the core and island.

There are two important remarks to make. The first concerns the fitting of the curves representing the beam losses recorded by the diamond detectors. As seen in Fig.~\ref{fig:fast_BLMs} (top), the spikes representing the losses are not symmetric with respect to their maximum value, which is an artefact introduced by the electronics. For this reason, only the rising part of the signal is used in our analyses, but this does not introduce any bias in the proposed method. The second remark concerns the use of the data measured by the diamond detectors in SS15 and SS16, as only the data set from the first detector can be exploited to extract information about the horizontal beam emittances. Indeed, the lack of magnetic field in the straight section where the dummy septum blade is installed ensures that the proposed method can be properly applied, whereas in the case of the detector in SS16, we cannot exclude that the magnetic field of the extraction septum (or its fringe field close to the blade, but still in the circulating beam aperture) affects the dynamics of the secondaries so as to introduce a bias in the measured data recorded by the diamond detector. 
\section{Simulation of the performance of the diamond detector} \label{sec:sim}
\subsection{Generalities on the diamond detector}
The diamond detector is produced by CIVIDEC~\cite{ref:pCVD} and consists of a polycrystalline diamond sensor (pCVD) with a size of~\SI{10}{mm} $\times$ \SI{10}{mm} and thickness of \SI{500}{\micro m}. Diamond is a semiconductor, and particles traversing the detector generate electrons and holes in pairs, which drift in the externally applied electric field and induce a current signal collected by a preamplifier. The nominal bias voltage is \SI{500}{V} and the signal is amplified by \SI{40}{dB}, and after about \SI{200}{m} of cable is read by a digitiser. The interest in this type of detector is that it has an excellent time response and can resolve the time structure of the beam losses with a resolution at the level of \SI{}{ns}~\cite{stein:ipac16-mopmr031,PhysRevAccelBeams.23.044802,PhysRevAccelBeams.23.124501}.

\subsection{Results of numerical simulations}
\begin{figure}[!htb]
   \centering
   \includegraphics[trim=0truemm 2truemm 0truemm 2truemm,width=0.9\textwidth,clip=]{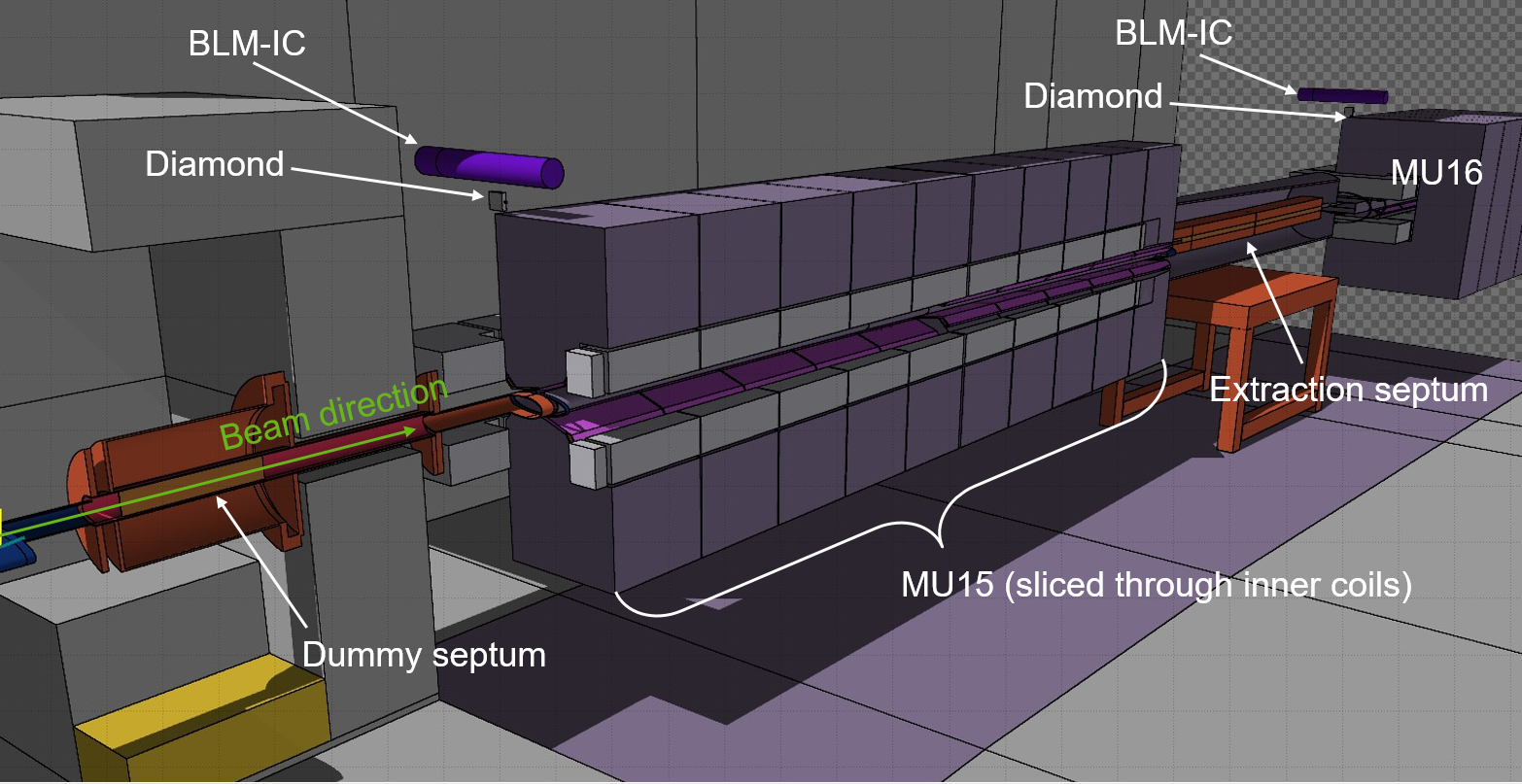}
   \caption{Cross section of selected elements included in the model used in the FLUKA simulations. The element downstream of MU15 are not relevant for the studies presented here.}
   \label{fig:geometry}
\end{figure}

To investigate the behaviour of the diamond detector and characterise its output, a detailed model of the region around the dummy septum was created in FLUKA~\cite{FLUKA:2015, FLUKA:2022}. Using Python-based software LineBuilder~\cite{LineBuilder}, accelerator elements and components were placed along a curved trajectory, based on the geometry of the PS ring calculated using the MAD-X code~\cite{madx}. Elements of the geometry description include the vacuum tank containing the dummy septum, concrete shielding around the dummy septum, and the main combined function (CF) magnet (MU15) downstream of the dummy septum, with the pCVD diamond detector housed above it. The geometry of the simulation can be seen in Fig.~\ref{fig:geometry}, including cross sections of relevant elements.

3D field maps generated in ANSYS\textsuperscript{\textregistered} were implemented for the CF magnet. Tracking of on-momentum, on-reference orbit protons was performed to tune the magnetic field strength so that the particles received the nominal \SI{30}{mrad} kick in the CF magnet. 

\begin{figure}[!htb]
   \centering
   \includegraphics[trim=0truemm 2truemm 0truemm 2truemm,width=0.8\textwidth]{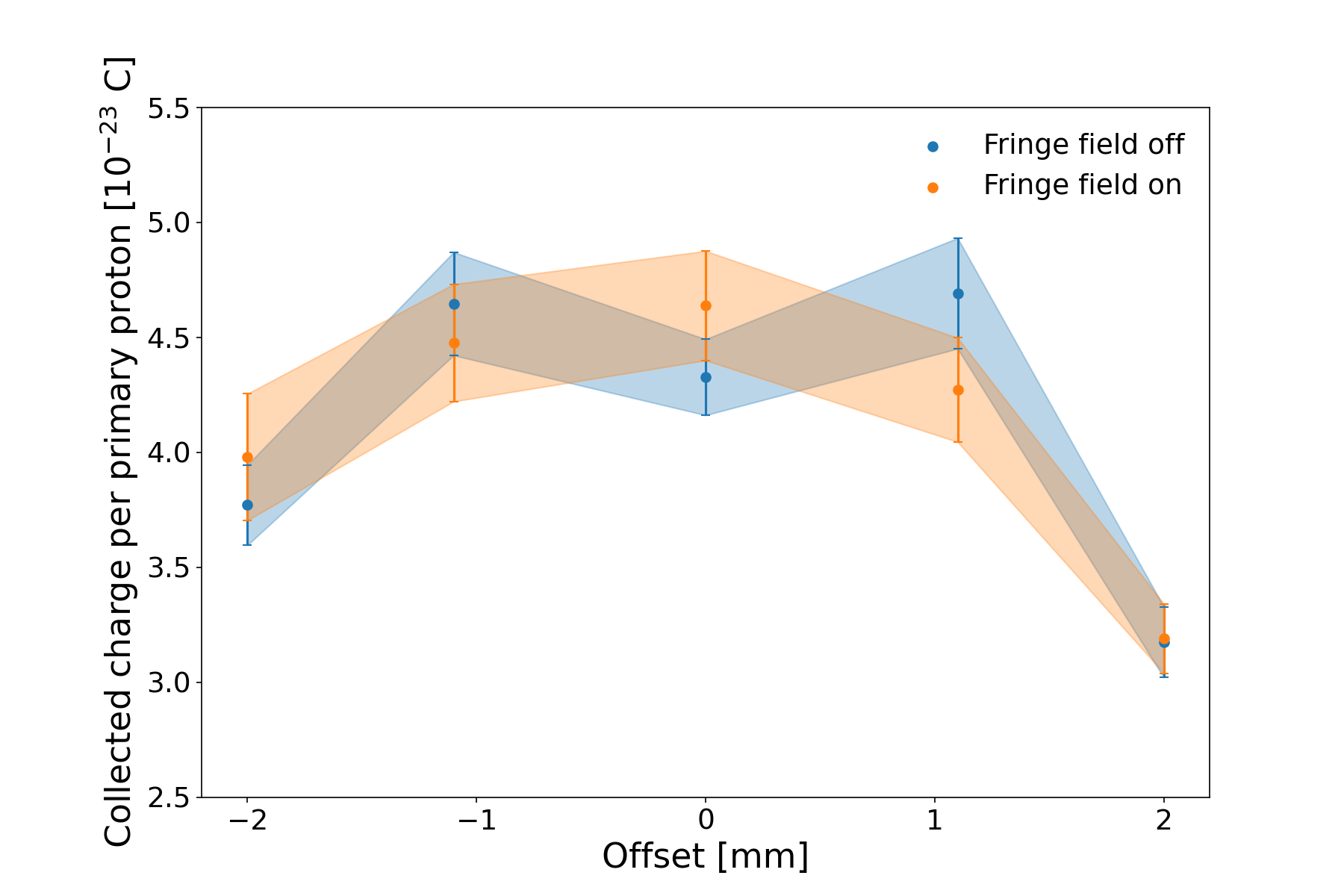}
   \caption{Charge collected by the diamond detector per primary proton, as a function of impacting particle's position on the dummy septum blade, in the presence and without fringe field. Typical difference between the case with and without the fringe field is of the order of $9$\%.}
   \label{fig:fringeFields}
\end{figure}

An investigation was carried out to determine the impact of the fringe fields protruding from the CF magnet between the dummy septum and the diamond detector. Simulations of protons impinging at different impact points on the \SI{4.2}{mm} wide front face of the dummy septum blade were performed with the fringe field present and removed. 
Figure~\ref{fig:fringeFields} shows a comparison of the charge collected per primary proton by the diamond detector as a function of the horizontal position of the proton in both cases. As expected, a small drop in the detector signal was observed at the outer edges of the septum due to less downstream material for a developing shower to propagate through. It was also seen that the two sets of simulations are statistically compatible, so subsequent simulations were performed without a fringe field.

\begin{figure}[!htb]
   \centering
   \includegraphics[trim=0truemm 2truemm 0truemm 2truemm,width=.9\textwidth]{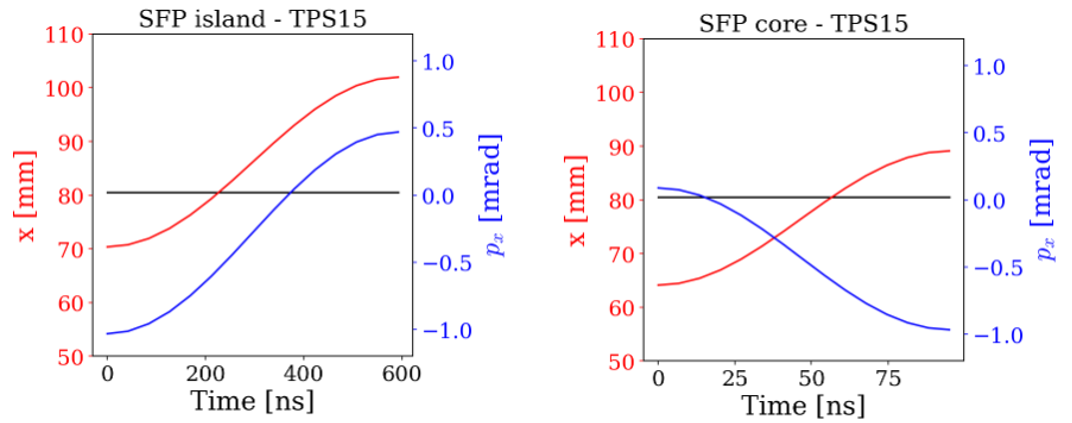}
   \caption{Beam displacement (red) and angle (blue) variation at the dummy septum as a function of time for the core (right) and the island (left).}
   \label{fig:riseTimes}
\end{figure}

To perform a simulation with a realistic distribution of the beam inside the island, multiturn tracking was performed using the PTC tracking library~\cite{Schmidt:573082,Forest:ICAP06-MOM1MP02} in MAD-X. In this case, the details of the splitting process were not simulated, and the beam distribution in the island is generated as a double Gaussian in the normalised horizontal phase space, and the coordinates are then transformed to the physical phase space, with the distribution being centred at the stable fixed point (SFP) inside the extracted island. The initial distribution, composed of $6 \times 10^4$ particles, is tracked for $2048$ turns, causing complete filamentation. Typically, about $3 \times 10^4$ particles are retained after the filamentation stage, as the others are not inside the island. The retained distribution within the island was used to produce initial conditions upstream of the dummy septum for the FLUKA simulations and was horizontally displaced to represent the increasing strength of the extraction kickers as a function of time. To simulate the extraction process, the horizontal position and angular offset of each primary proton were determined at the specific time considered. Using the TWISS module in MAD-X, it was possible to determine the time evolution of the island SFP angle and position from knowledge of the induced kicker displacement as reported in Eq.~\eqref{eq:risetime}. These values were then interpolated during sampling and applied to each primary proton accordingly. Displacements and deflections of the SFP position upstream of the dummy septum during extraction can be seen in Fig.~\ref{fig:riseTimes} (for the core, right, and the island, left) as a function of time.

Using this approach, we performed a simulation to compare the profile of the proton losses as a function of time with the profile given by the energy deposited in the diamond detector by secondary particles generated by the interaction of the beam with the copper blade of the dummy septum. The profile of the energy deposited in the diamond detector simulated with FLUKA was scaled assuming a conversion factor of \SI{4}{mV} per interacting minimum ionising particle (MIP)~\cite{CableAttenuation}. Furthermore, it was assumed that the average energy deposited in the active material of the diamond detector on a range of secondary particles species (electron, pion, proton) is \SI{0.3205}{MeV} per MIP, calculated by a FLUKA simulation. The attenuation of the signal along the $\approx \SI{200}{m}$ long cables has been assumed to be in the range of $20-30\%$~\cite{CableAttenuation}. 

\begin{figure}[!htb]
   \centering
   \includegraphics[trim=0truemm 2truemm 0truemm 2truemm,width=0.8\textwidth]{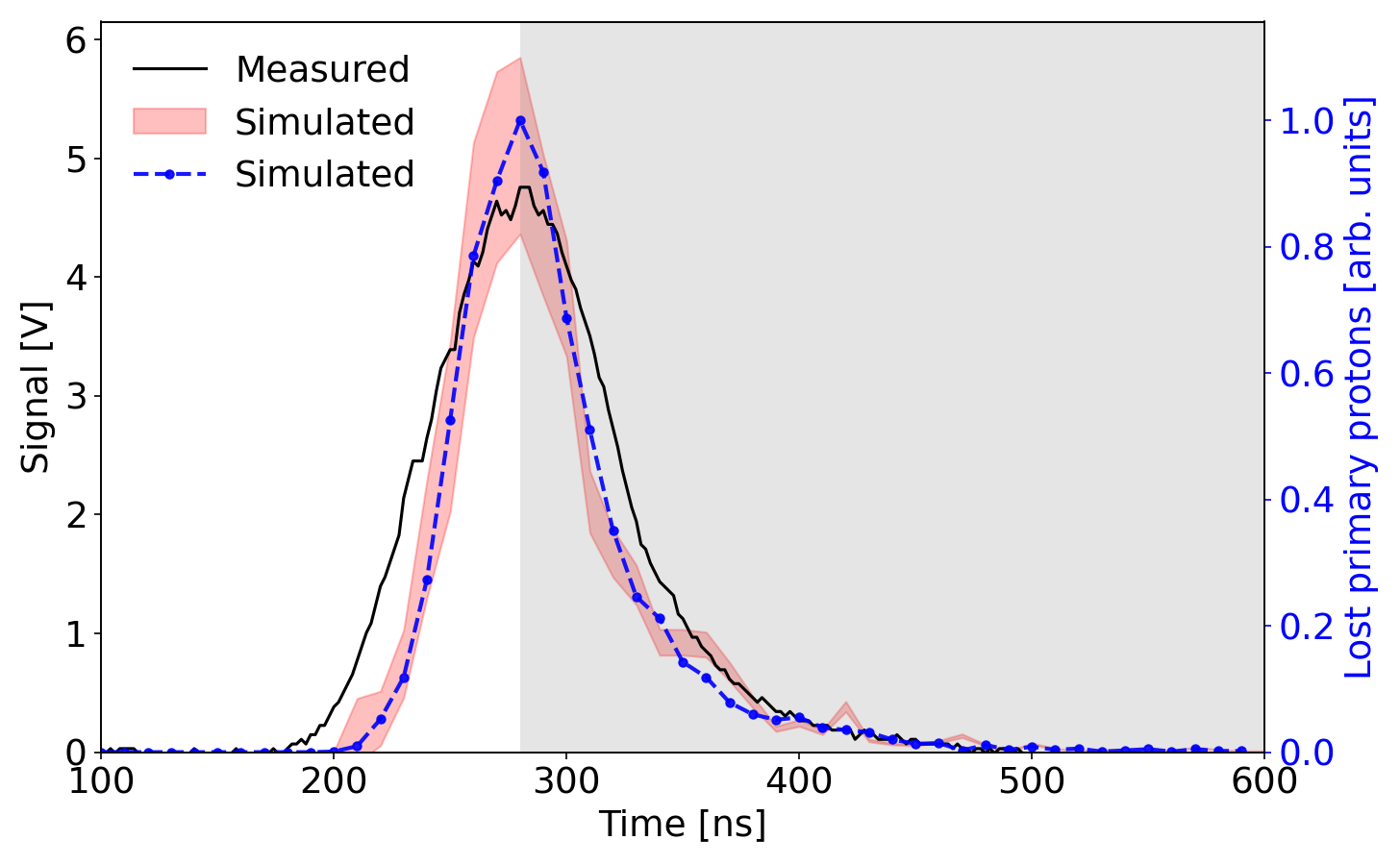}
   \caption{Comparison of the simulated signal profile (red band, left axis) with the number of simulated lost protons (blue line, right axis) as a function of time for the extraction of the island for the diamond detector in SS15. These data are compared with the measured beam losses recorded on the $15^{th}$ of November 2017 at 21:34:54 with a beam intensity of $1.85 \times 10^{13}$~protons (black line, left axis). The band of the simulated signal includes the cable attenuation (see next section) and the statistical uncertainty of the simulation. The lost proton profile is computed from simulation (refer to the text for more detail). In the studies reported in the rest of the paper, only the raising edge of the loss profile is taken into account, i.e. the portion of the signal in the shaded area is excluded.}
   \label{fig:losses}
\end{figure}

Figure~\ref{fig:losses} shows excellent agreement between the profile of the lost primaries (blue line), which are assumed to be the protons impacting the dummy septum blade considered as a perfect absorber, and the simulated signal seen in the detector (red band), where the band corresponds to the cable attenuation (see next section) and the statistical uncertainty of the simulation. This observation confirms that the time profile of the signal provided by the diamond detector precisely follows the beam losses during the rise time of the extraction kickers. A comparison with the measured data obtained with a beam intensity of $1.85\times 10^{13}$~protons is also seen in Fig.~\ref{fig:losses} (black line). The slight difference in the rising edge of the measured and simulated loss profile may be related to a slight difference in the rise time of the kickers used in the simulation with respect to the actual one, while the difference in the falling edge is expected to be associated with the electronics readout. Note that in the data analyses presented in the next sections, only the rising edge of the signal is used, i.e. the portion of the signal in the shaded area is excluded.

In the numerical simulations performed, the angle of the closed orbit was neglected. Therefore, additional simulations were carried out including a non-zero closed orbit angle (on the order of $\pm \SI{1}{mrad}$, which corresponds to the typical uncertainty of this parameter). They show that the raising edges of the signal readout by the diamond detector are largely independent of the value of the angle of the closed orbit at the location of the dummy septum. This indicates that the assumptions made should not affect the accuracy of the emittance reconstruction.  
\section{Experimental results} \label{sec:emitt}
\subsection{Validation of the model assumptions}
The first step is to scrutinise the assumptions used to build the model describing the beam losses, which provide a hint on how to extract the numerical value of the transverse rms beam size of the beam using the loss data. According to Eq.~\eqref{eq:losses1}, the measured beam losses $\ell_\mathrm{m}(t)$ should be well represented by a Gaussian. The fit of the numerical data confirmed this hypothesis since all are of excellent quality (see, for instance, the fit shown in Fig.~\ref{fig:fast_BLMs}, bottom). This implicitly shows that the approximation used to represent the integral and the function $\mu(t)$ is correct. Furthermore, it also indicates that the diamond detector does not introduce any sizeable non-linear distortion of the theoretical signal proportional to the number of protons lost on the dummy septum blade; otherwise, the expected Gaussian time profile would be affected.

\begin{figure}[!htb]
   \centering
   \includegraphics[trim=0truemm 2truemm 0truemm 2truemm,width=0.8\textwidth, clip=]{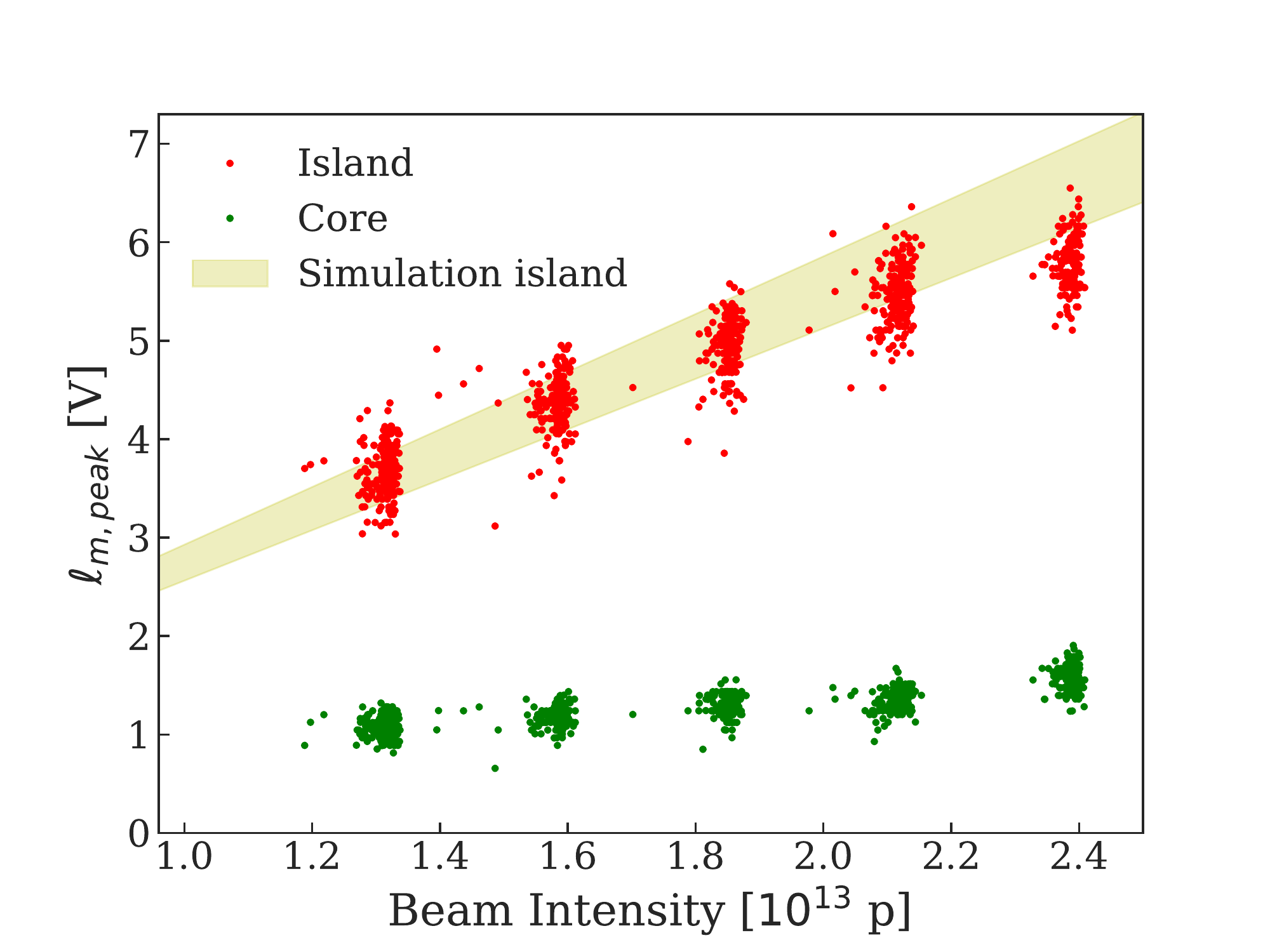}
   \caption{Peak value of the measured losses as a function of the beam intensity for the core and island. Data from numerical simulations carried out with FLUKA are also shown, including the uncertainty on the cable attenuation (band).}
   \label{fig:signal}
\end{figure}

Furthermore, the linearity of the detector has been probed in a more direct way by inspecting the relationship between $\ell_\mathrm{m, peak}$, the peak of the measured losses, and the total beam intensity. This is shown in Fig.~\ref{fig:signal}, where the data have been split to distinguish between the core and the island.

Some slight sign of non-linear behaviour is seen for the dataset corresponding to the highest beam intensity probed during the experimental session. To make this observation more quantitative, the data for the island were fitted with a straight line but using only the first four groups of beam intensities. The fitted line provides an estimate $\ell_\mathrm{m, peak}=(6.2\pm 0.3) \SI{}{V}$ for the case corresponding to the average intensity of the last group of data. On the other hand, the data corresponding to the last intensity range give $\ell_\mathrm{m, peak}=(5.8\pm 0.3) \SI{}{V}$. This estimate differs by about $6\%$ with respect to the extrapolation based on the linear fit, which can be considered as an estimate of the deviation from the linear behaviour of the diamond detector. All in all, such a deviation, although not zero, is very small and, as will be shown in the next sections, it has a negligible impact on the data analyses. 
\subsection{Reconstruction of the beam emittance in the linear and nonlinear cases}
According to the method outlined in the previous section, the reconstruction of the emittance for the core is relatively straightforward. The time evolution of the beam losses is fitted with a Gaussian model, and the beam sigma is obtained by applying Eq.~\eqref{eq:sigma}, from which the emittance is obtained by using
\begin{equation}
    \epsilon_\mathrm{H}^\ast = \beta_\mathrm{rel} \gamma_\mathrm{rel} \frac{\sigma^2-\sigma^2_\mathrm{p} D_\mathrm{H}^2}{\beta_\mathrm{H}} \, , 
    \label{eq:emittance_lin}
\end{equation}
where $\beta_\mathrm{rel}, \gamma_\mathrm{rel}$ are the relativistic parameters, $\beta_\mathrm{H}, D_\mathrm{H}$ are the horizontal beta and dispersion functions, respectively, and $\sigma_\mathrm{p}=0.5\times 10^{-3}$ is the rms beam momentum spread. This standard approach is feasible because the dynamics of the core is essentially linear. The nominal optical parameters have been used, since optics measurements carried out in the past confirmed that the maximum beta-beating is smaller than a few percent~\cite{Hernalsteens:IPAC13-WEPEA054,Skowro:IPAC19-MOPTS102} and therefore completely negligible.

\begin{figure}[!htb]
   \centering
   \includegraphics[trim=4truemm 5truemm 10truemm 10truemm,width=0.49\columnwidth,clip=]{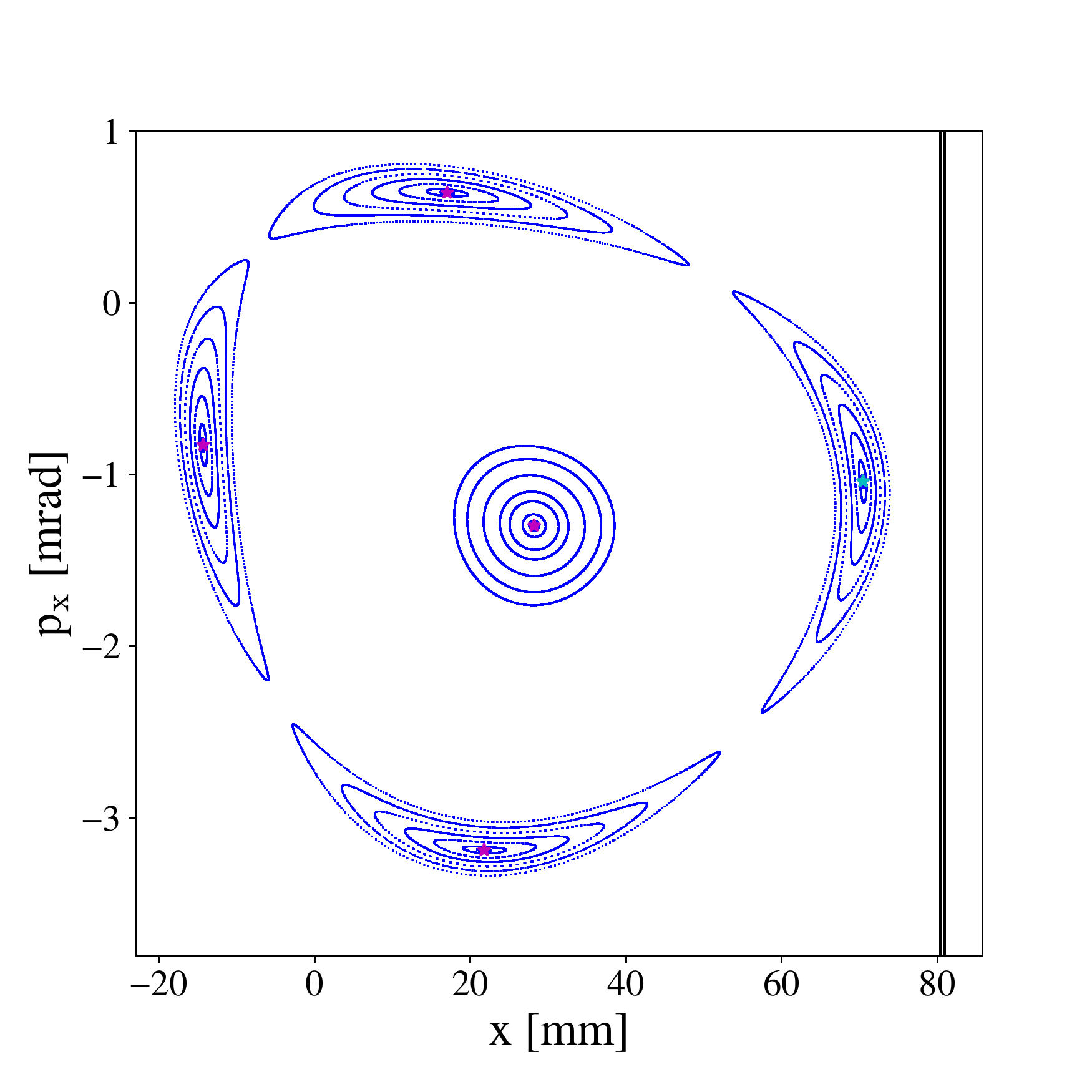}
   \includegraphics[trim=4truemm 5truemm 10truemm 10truemm,width=0.49\columnwidth,clip=]{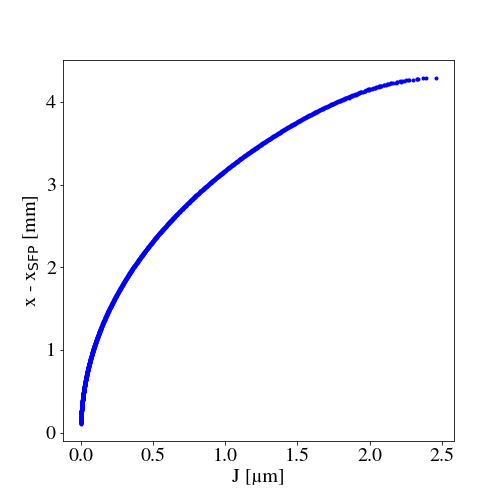}
   \caption{Left: Phase-space representation at SS15 of the core and islands. The stable fixed points (SFP) are highlighted with a magenta dot, except for the rightmost, which represents the extracted island, that is represented as a cyan dot. The vertical line indicates the position of the dummy septum. Right: relationship between the horizontal position inside the extracted island referred to the position of the SFP and the corresponding action value. For each value of $x-x_\mathrm{SFP}$ the orbit of this initial condition is computed and its action estimated.}
   \label{fig:PhaseSpace_SS15_Action_SFP_position}
\end{figure}

The situation is completely different when the beamlets are considered, since the motion inside the stable islands is intrinsically nonlinear, and this imposes a different approach to the postprocessing of the measured data. The situation can be seen in Fig.~\ref{fig:PhaseSpace_SS15_Action_SFP_position} (left) where the horizontal phase-space portrait is shown at the location of the dummy septum. 
The four islands corresponding to the four beamlets are visible, and the points at their centre are the SFP: the one with the largest $x$ coordinate represents the island that is used for extraction from the PS ring. While the dynamics around the origin of the phase space is very close to circles, i.e. the standard situation in the case of linear motion, around the fixed points the closed curves are represented by tilted and deformed ellipses. The deformation increases with distance from the fixed points. This means that the linear optical parameters around the fixed points differ from those of the nominal ring\footnote{This observation is at the heart of a proposed novel approach to transition crossing in a circular accelerator~\cite{GiovannozziEPJPlusTransition}.}. Their values can be computed by using the full non-linear model of the PS ring together with MAD-X~\cite{madx} and PTC~\cite{Schmidt:573082,Forest:ICAP06-MOM1MP02,Forest:ICAP06-WEPPP04}, which allows the linear optics around an arbitrary stable fixed point to be evaluated, even when the stable fixed point is different from the origin, i.e. the standard closed orbit. 

We recall that the non-linear model of the PS ring is based on beam measurements, in which the tunes as a function of momentum offset are determined and these functions are fitted by introducing thin-lens elements of sextupolar and octupolar type into each of the half-units of the main magnets of the PS~\cite{Giovannozzi:PAC03-RPAG012}, so that their four strengths can be used to reproduce the measured curves.

\begin{figure}[!htb]
   \centering
   \includegraphics[trim=4truemm 0truemm 10truemm 8truemm,width=0.49\columnwidth,clip=]{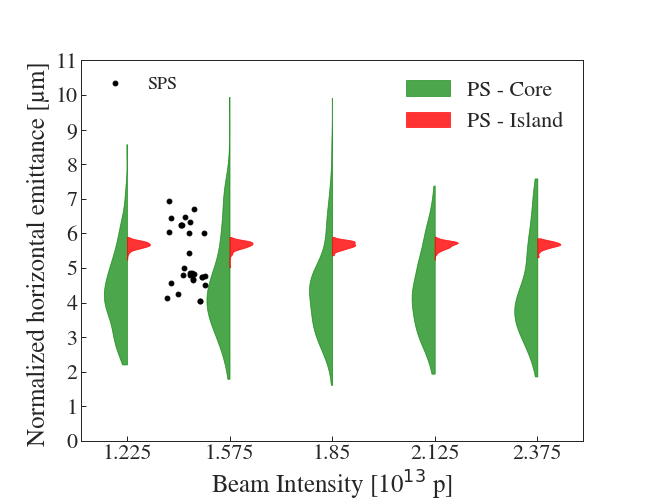}
   \includegraphics[trim=4truemm 0truemm 10truemm 10truemm,width=0.49\columnwidth,clip=]{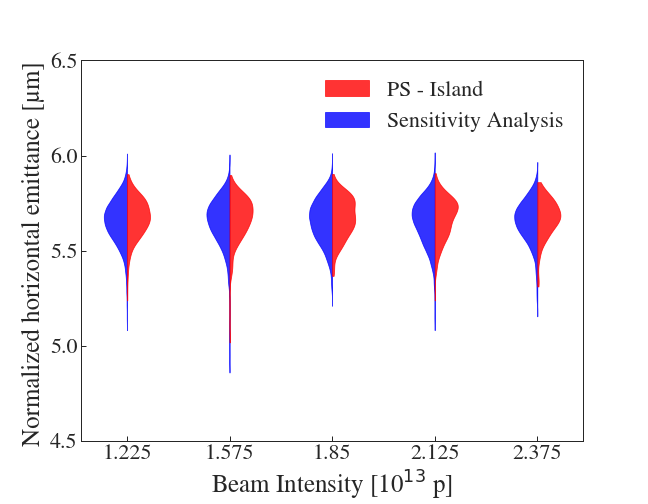}
   \caption{Left: Distribution of the normalised beam emittance for the core and the island as a function of the total beam intensity. No visible trend is observed. The reconstruction of the normalised beam emittances is performed according to the method described in the text. For the sake of comparison, emittance measurements performed in the SPS after injection are also shown. Right: comparison between the reconstructed normalised emittance of the island and the emittance distribution obtained from the sensitivity analysis on the nonlinear PS model.}
   \label{fig:SensitivityAnalysis}
\end{figure}

Although the optical beam parameters around the stable fixed points are known, they cannot be used in combination with Eq.~\eqref{eq:emittance_lin} to derive a value of the beam emittance due to non-linear dynamics. Indeed, it is necessary to go back to the definition of the beam emittance in terms of action, namely
\begin{equation}
    \epsilon = \langle J \rangle = \int_0^{J_\mathrm{sep}} \mathrm{d} J J \rho(J) \, , 
    \label{eq:emittance_nlin}
\end{equation}
where $J_\mathrm{sep}$ represents the maximum value of the action reached at the separatrix that surrounds each of the islands and $\rho(J)$ represents the distribution of the beam within the island as a function of the action. The beam distribution inside the beamlet can be extracted from the beam losses using the approach described in the previous section. This beam distribution is represented by a Gaussian function of the physical coordinates, the sigma of which is given by $\sqrt{\sigma^2-\sigma^2_\mathrm{p} D_\mathrm{H}^2}$. However, there is still to determine the relationship between the measured beam distribution, which is obtained in physical coordinates, and the action. To this aim, the non-linear model can be used, as, by selecting initial conditions of the form $(x_\mathrm{SFP}+x_\mathrm{scan},p_{x,\mathrm{SFP}})$ with $x_\mathrm{scan}$ ranging from zero to the separatrix, their orbits can be computed by direct tracking, and the corresponding action can be obtained by applying its definition, namely
\begin{equation}
    2\pi J = A \quad \text{where} \quad A = \text{surface enclosed by the orbit} \, .
\end{equation}
The curve shown in Fig.~\ref{fig:SensitivityAnalysis} (right) is obtained in this way and represents the link between the physical coordinate and the action, thus allowing the evaluation of Eq.~\eqref{eq:emittance_nlin}. The result of the analysis discussed here for both the core and the island is shown in Fig.~\ref{fig:SensitivityAnalysis} (left), where the distributions of the normalised emittance for the core and the island are shown for the whole data set made of $999$ beam measurements, plotted as a function of the total beam intensity.

The first observation is that the horizontal beam emittance does not depend on the total beam intensity. This is not surprising, as the process that induces beam splitting is essentially independent of the intensity of the beam (some intensity-dependent effects have been observed in the beam splitting process~\cite{PhysRevSTAB.16.051001}, but they are irrelevant in this context). This also confirms that the slight sign of the non-linear response of the diamond detector to beam losses does not affect the value of the reconstructed beam emittance. 

The second observation is that, while the emittance values of the core are typically smaller than those of the island (by comparison of the most probable value of the various distributions), the spread of their distribution is very different. This is a very interesting observation that could not be made before this method was attempted, which shows its relevance and usefulness. It will be considered for future investigations, and it is speculated that the large spread of the core emittance distribution is due to the interaction between the transverse damper, which is used to provide better control of the trapping process~\cite{PhysRevAccelBeams.20.061001}, and the tuning stability of the PS ring~\cite{Giovannozzi:IPAC17-WEPIK088}. 

Some emittance measurements performed at the SPS are also shown for the sake of comparison. We recall that the beam injected into the SPS has lost the special structure it had in the PS prior to extraction, namely, the distinction between the core and the island. Therefore, a standard emittance measurement can be performed using a flying wire and records an emittance value that depends on that of the core and island. In this respect, the scattered black points representing the SPS measurements are clearly compatible with the reconstructed emittance values at the PS. It is also worth mentioning that the SPS emittance measurements were not performed during the same data collection period during which the PS data were collected. Therefore, the observed agreement between the PS results and the SPS data is considered quite good.
\subsection{Sensitivity analysis of the reconstruction method for the nonlinear case}
From the considerations made in the previous section, it is clear that the reconstruction of the emittance of the island is heavily dependent on the nonlinear model of the PS. For this reason, a sensitivity analysis was performed. 

\begin{figure}[!htb]
   \centering
   \includegraphics[trim=0truemm 2truemm 0truemm 2truemm,width=0.49\columnwidth,clip=]{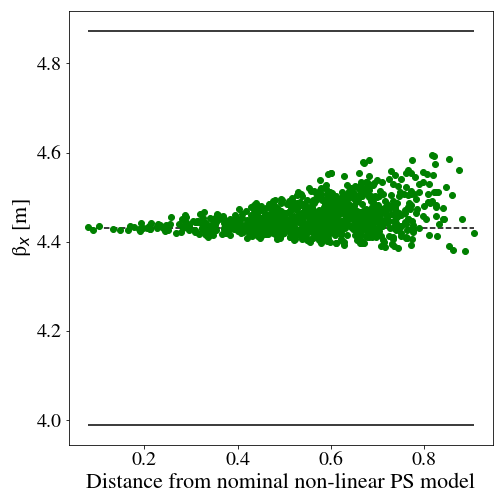}
   \includegraphics[trim=0truemm 2truemm 0truemm 2truemm,width=0.49\columnwidth,clip=]{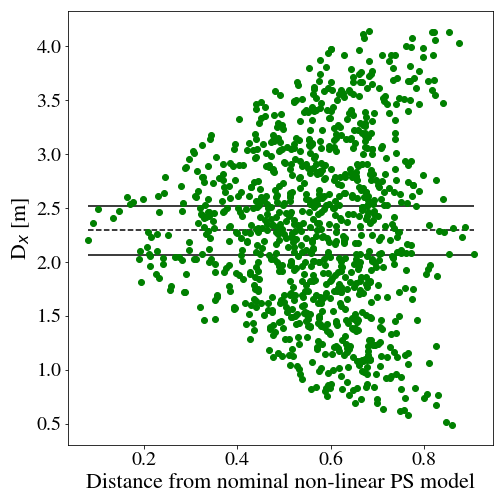}
   \includegraphics[trim=0truemm 2truemm 0truemm 2truemm,width=0.49\columnwidth,clip=]{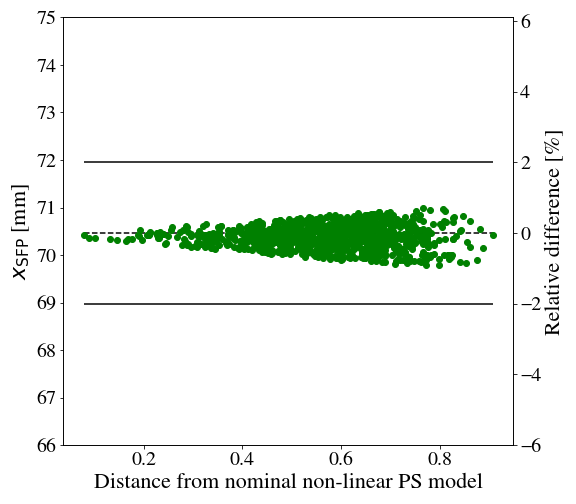}
   \caption{Horizontal beta-function (top-left), dispersion function (top-right), and position of the SFP and the dummy septum location (bottom) as a function of the euclidean distance of the sextupolar and octupolar components of the synthetic non-linear PS models from the nominal one. The horizontal continuous lines in the top plots represent a variation of $\pm 10\%$ around the nominal value of the parameter. The tolerance for the SFP position is set to $\pm$ \SI{1.5}{\milli\meter} around the nominal value.}
   \label{fig:sensitivity_optics}
\end{figure}

The non-linear sextupolar and octupolar components, assigned to the half-units that make up each of the main PS magnets, have been varied over an interval corresponding to $\pm 2\%$ of their value in the nominal non-linear PS model. The non-linear components of such a 4D hypercube have been used to generate $10^3$ synthetic nonlinear PS models that have been used to evaluate the optical parameters around the fixed points, as well as the position of the fixed point that represents the extracted beamlet at the location of the dummy septum. The result of these calculations is shown in Fig.~\ref{fig:sensitivity_optics}.
The parameters depicted in the plots are shown as functions of the Euclidean distance of the corresponding sextupolar and octupolar components from the values of the components in the nominal non-linear PS model. 

The variation of $\beta_x$ that describes the motion around the SFP (top left graph) is biased toward values larger than the nominal one, but all reported values are within the interval of $\pm 10\%$, which is represented by continuous horizontal lines. However, the horizontal dispersion (top right graph) features a substantial variation around the nominal value, generating a large number of cases well outside of the range $\pm 10\%$. In the case of the position of SFP, shown in Fig.~\ref{fig:sensitivity_optics} (bottom plot), the calculated value is shown together with the acceptance interval corresponding to $\pm$~\SI{1.5}{\milli\meter}. Acceptance intervals have been established on the basis that knowledge of the optical parameters of the PS ring is typically at the level of $10\%$ and that a variation of the SFP position beyond $\pm$~\SI{1.5}{\milli\meter} is easily detected at the level of measured extraction conditions. Cases that fall within the acceptance intervals, which are $204$ of an initial number of $10^3$ configurations, have been used to reconstruct the emittance values from the measured data using the method proposed earlier. The results of the sensitivity analysis and those using the nominal nonlinear PS model are compared in Fig.~\ref{fig:SensitivityAnalysis} (right). The mean, the most probable value, and the spread of the emittance distributions for the two cases, namely the nominal nonlinear model and its variations, are extremely close to each other, and this provides clear evidence that the nominal nonlinear model is reliable for reconstructing the island emittance.
\section{Considerations on the performance of the MTE beams} \label{sec:perf}
In this section, we discuss how our new emittance measurement method can open new paths of performance optimisation for the MTE beam.
\subsection{Emittance reduction}
Among the several goals and merits of MTE, one is to provide a reduction of the initial horizontal beam emittance. Thanks to the proposed method, it is possible to quantify the emittance reduction for both island and core, by defining the emittance reduction factor as:
\begin{equation}\label{Reduction_Factor}
    F = \frac{\langle \epsilon^*\rangle}{\langle \epsilon_\mathrm{H}^*\rangle} \,
\end{equation}
where $\langle \epsilon_\mathrm{H}^* \rangle$ and $\langle \epsilon^* \rangle $ are the average horizontal beam emittance before splitting and the corresponding value after splitting, respectively. 

Taking into account the typical normalised emittance values of the beam before splitting, which are around 10-\SI{12}{\micro m} and using the average values of the distributions shown in Fig.~\ref{fig:SensitivityAnalysis} (for the island, the values corresponding to the nominal case are used), one obtains $F_\mathrm{core}=0.3-0.4$ and $F_\mathrm{island}=0.5-0.6$. In principle, the theoretical target of beam splitting is to share equally the initial emittance between the core and the islands, hence $F_\mathrm{th}=0.2$. Larger values of $F$ are explained by the non-linear dynamics that introduces emittance blow-up, as the beam emittance is not preserved by the non-linear effects. This fact is observed for both the core and the islands, although the largest effect is observed for the latter. It is also clear that in the islands motion is strongly non-linear; hence larger $F$-values are to be expected. Therefore, the new diagnostic means provided by the technique described above will be an essential tool for studying and optimisation of the emittance-sharing process.

\subsection{Injection matching to the SPS}
Injection matching to the SPS is particularly challenging for the MTE beam. Indeed, in the transfer line joining the PS to the SPS, the beam is made of the extracted island (with a length corresponding to four times the circumference of the PS) followed by the core (with a length corresponding to one time the circumference). The two parts of the total beam feature different optical parameters (Twiss and dispersion) at the PS extraction point, and they will experience the same magnetic field in the transfer line; hence, they will have different optical parameters at SPS injection. This creates different betatron and dispersion mismatches for the two parts, which eventually might turn into different emittance growth. To optimise the SPS injection performance, in particular, to avoid losses related to emittance growth, a criterion for setting the optical settings of the transfer line should be fixed. A possibility could be to set the optics of the transfer line so that either the core or the island are matched at SPS injection, as was tried during the initial stages of the MTE operations. However, this is not necessarily the best approach. 
Let us first briefly review the theory of emittance growth as a result of injection mismatch. Two phenomena are possible, namely, betatronic mismatch and dispersion mismatch~\cite{edwards_syphers_1993}, and, in general, a combination of the two occurs. Let us indicate with $\alpha_\mathrm{u}, \beta_\mathrm{u}, \gamma_\mathrm{u}, D_\mathrm{u}, D'_\mathrm{u}$ the injection parameters for the beam $u$, where $u=i, c$ for the island or the core, respectively, and let us indicate with $\alpha^\mathrm{inj}, \beta^\mathrm{inj}, \gamma^\mathrm{inj}, D^\mathrm{inj}, {D'\,}^\mathrm{inj}$. The emittance after filamentation $\epsilon_\mathrm{u}^\mathrm{af}$ due to the betatron mismatch is given by
\begin{equation}
    \epsilon_\mathrm{u, \beta}^\mathrm{af} = \frac{\epsilon_\mathrm{u}}{2} \left ( \beta_\mathrm{u} \gamma^\mathrm{inj} -2 \alpha_\mathrm{u} \alpha^\mathrm{inj} + \gamma_\mathrm{u} \beta^\mathrm{inj} \right ) \quad \mathrm{u} = \mathrm{i}, \mathrm{c} \, , 
    \label{eq:betamis}
\end{equation}
and in the case of a dispersion mismatch, one has
\begin{equation}
    \epsilon_\mathrm{u, D}^\mathrm{af} = \epsilon_\mathrm{u} + \frac{\Delta D_\mathrm{u}^2}{2 \beta^\mathrm{inj}} \sigma_\mathrm{p}^2 \quad \mathrm{u} = \mathrm{i}, \mathrm{c} \, , 
    \label{eq:dispmis}
\end{equation}
where
\begin{equation*}
  \Delta D_\mathrm{u}^2 = \Delta D^2 + \left ( \beta^\mathrm{inj} \Delta D' + \alpha^\mathrm{inj} \Delta D \right)^2 \quad \text{with} \quad \Delta D = D^\mathrm{inj}-D_\mathrm{u} \quad \Delta D'= {D'\,}^\mathrm{inj}- D'_\mathrm{u} \quad \mathrm{u} = \mathrm{i}, \mathrm{c} \, . 
\end{equation*}

Different optical parameters at PS extraction for the core and island generate, for the same transfer line configuration, different optical parameters for injection into the SPS. It is clearly impossible to achieve simultaneous matching of the optical parameters of the core and island at SPS injection with the nominal SPS parameters. Therefore, the best compromise consists in defining a function to be optimised by selecting the appropriate optical functions of the transfer line. The most appropriate criterion is to minimise the beam size after filamentation to minimise the possible beam losses due to aperture limits. Therefore, the optimal configuration of the transfer line corresponds to minimising the following function.
\begin{equation}
   \mathcal{W} = w_\mathrm{i} \left (\sqrt{\beta_\mathrm{i} \epsilon_\mathrm{i, \beta}^\mathrm{af}} +  \sqrt{\beta_\mathrm{i} \epsilon_\mathrm{i, D}^\mathrm{af}} \right ) + \left (\sqrt{\beta_\mathrm{c} \epsilon_\mathrm{c, \beta}^\mathrm{af}} +  \sqrt{\beta_\mathrm{c} \epsilon_\mathrm{c, D}^\mathrm{af}} \right ) \, .
\end{equation}

The quantity $w_\mathrm{i}$ is an appropriate weight that takes into account the sharing of intensity between the island and the core, so that the minimisation of $\mathcal{W}$ turns out to act mainly on the emittance of the most intense part of the beam. In the optimal case, the island contains four times more beam than the core and $w_\mathrm{i}=4$, but the actual value could be adapted to specific needs. The definition of emittance after filamentation from Eqs.~\eqref{eq:betamis} and~\eqref{eq:dispmis} is based on the knowledge of the beam emittance of the core and island at extraction from the PS. It should be noted that the developed diagnostic method provides key information for the optimisation of the transfer line optics for the MTE beam, which might be used in future studies.
\section{Conclusions} \label{sec:conc}
In this paper, a nondestructive method to measure the horizontal beam emittance of the transversely split proton beam produced by the CERN PS ring has been presented and discussed in detail. The proposed technique is based on the measurement of beam losses that occur during beam extraction, which are due to interactions with the copper blade of a special device called a dummy septum. These losses, recorded by means of a diamond detector that allows for fast data acquisition, on the order of nanoseconds, can be fitted by a Gaussian model to recover the information on the transverse beam profile, which provides the link to the beam emittance. It is worth stressing that while this method can be rather straightforwardly applied to the determination of the emittance of the core beam, a special treatment is needed for the beam trapped in the stable islands. Indeed, the intrinsic non-linear motion inside the islands imply a number of special treatments of the collected data, including the use of numerical simulations of the non-linear model of the beam dynamics. These treatments represent one of the novelties of the proposed approach. 

The method has been successfully applied to measured data collected in 2017 during a dedicated experimental session. However, a key aspect of the proposed technique is that it could be used during routine operation for diagnostic purposes. Indeed, this approach has allowed the measurement of the horizontal beam emittance of both core and island beams for the first time, and it has opened up the possibility for future studies and optimisations of the performance of the split beam in the PS ring, which have been presented and discussed in the last part of the paper.  
\section*{Acknowledgments}
We thank our colleagues E.~Calvo Giraldo, E.~Effinger, R.~Garcia Alia, D.~Lampridis, B.~Salvachua Ferrando and C.~Zamantzas, for helpful discussions and support. We would also like to warmly acknowledge F.~Capoani for several discussions on the action distribution inside the islands.


\bibliographystyle{unsrt}
\bibliography{mybibliography}

\providecommand{\noopsort}[1]{}\providecommand{\singleletter}[1]{#1}%
\begin{thebibliography}{10}

\bibitem{Bovet:880676}
Claude Bovet, D~Fiander, L~Henny, A~Krusche, and G~Plass.
\newblock {The fast shaving ejection for beam transfer from the CPS to the CERN
  300 GeV machine}.
\newblock {\em IEEE Trans. Nucl. Sci.}, 20:438--41, 1973.

\bibitem{Gilardoni:EPAC08-THPC047}
S.S. Gilardoni and J.~Barranco.
\newblock {Studies of Losses During Continuous Transfer Extraction at the CERN
  proton Synchrotron}.
\newblock In {\em Proc. 11th European Particle Accelerator Conf. (EPAC'08)},
  pages 3083--3085. JACoW Publishing, Jun. 2008.

\bibitem{PhysRevSTAB.14.030101}
J.~Barranco~Garc\'{\i}a and S.~Gilardoni.
\newblock {Simulation and optimization of beam losses during continuous
  transfer extraction at the CERN Proton Synchrotron}.
\newblock {\em Phys. Rev. ST Accel. Beams}, 14:030101, 2011.

\bibitem{PhysRevLett.88.104801}
R.~Cappi and M.~Giovannozzi.
\newblock Novel method for multiturn extraction: Trapping charged particles in
  islands of phase space.
\newblock {\em Phys. Rev. Lett.}, 88:104801, 2002.

\bibitem{PhysRevSTAB.7.024001}
R.~Cappi and M.~Giovannozzi.
\newblock Multiturn extraction and injection by means of adiabatic capture in
  stable islands of phase space.
\newblock {\em Phys. Rev. ST Accel. Beams}, 7:024001, 2004.

\bibitem{Giovannozzi:987493}
M.~Giovannozzi, M.J. Barnes, O.E. Berrig, A.~Beuret, J.~Borburgh, P.~Bourquin,
  R.~Brown, J.-P. Burnet, F.~Caspers, J.-M. Cravero, T.~Dobers, T.~Fowler, S.S.
  Gilardoni, M.~Hourican, W.~Kalbreier, T.~Kroyer, F.~Di~Maio, M.~Martini,
  V.~Mertens, E.~Métral, K.D. Metzmacher, C.~Rossi, J.-P. Royer, L.~Sermeus,
  R.~Steerenberg, G.~Villiger, and T.~Zickler.
\newblock {\em {The {CERN} {PS} multi-turn extraction based on beam splittting
  in stable islands of transverse phase space: Design Report}}.
\newblock {CERN} Yellow Reports: Monographs. {CERN}, Geneva, 2006.

\bibitem{Bartosik:1463324}
H~Bartosik, D~Bodart, J~Borburgh, R~Brown, S~Damjanovic, S~Gilardoni,
  M~Giovannozzi, B~Goddard, C~Hernalsteens, M~Hourican, and M~Widorski.
\newblock {Proposal of a Dummy Septum to Mitigate ring irradiation for the
  {CERN} {PS} Multi-Turn Extraction}.
\newblock {\em Conf. Proc.}, C1205201:MOPPD059. 3 p, May 2012.

\bibitem{Bertone:1697680}
C~Bertone, J~Borburgh, D~Bodart, R~Brown, S~Burger, S~Damjanovic, P~Demarest,
  R~Fernandez~Ortega, JA~Ferreira~Somoza, D~Gerard, S~Gibson, S~Gilardoni,
  M~Giovannozzi, G~Le~Godec, C~Hernalsteens, M~Hourican, N~Jurado, J~M Lacroix,
  S~Mataguez, G~Métral, C~Pasquino, E~Perez-Duenas, S~Persichelli, B~Salvant,
  R~Steerenberg, and P~Van~Trappen.
\newblock {Studies and implementation of the {PS} dummy septum to mitigate
  irradiation of magnetic septum in straight section 16}.
\newblock Technical report, {CERN}, Geneva, Apr 2014.

\bibitem{Vadai:2702852}
M.~Vadai, A.~Alomainy, H.~Damerau, S.~Gilardoni, M.~Giovannozzi, and
  A.~Huschauer.
\newblock {Barrier bucket and transversely split beams for loss-free multi-turn
  extraction in synchrotrons}.
\newblock {\em EPL}, 128(1):14002. 7 p, 2019.

\bibitem{Vadai:IPAC19-MOPTS106}
M.~Vadai, A.~Alomainy, and H.~Damerau.
\newblock {Barrier Bucket Studies in the CERN PS}.
\newblock In {\em Proc. 10th Int. Particle Accelerator Conf. (IPAC'19)}, pages
  1128--1131. JACoW Publishing, May 2019.
\newblock https://doi.org/10.18429/JACoW-IPAC2019-MOPTS106.

\bibitem{Vadai:IPAC19-MOPTS107}
M.~Vadai, A.~Alomainy, H.~Damerau, S.~S. Gilardoni, M.~Giovannozzi, and
  A.~Huschauer.
\newblock {Beam Manipulations With Barrier Buckets in the CERN PS}.
\newblock In {\em Proc. 10th Int. Particle Accelerator Conf. (IPAC'19)}, pages
  1132--1135. JACoW Publishing, May 2019.
\newblock https://doi.org/10.18429/JACoW-IPAC2019-MOPTS107.

\bibitem{PhysRevAccelBeams.25.050101}
M.~Vadai, A.~Alomainy, H.~Damerau, M.~Giovannozzi, and A.~Huschauer.
\newblock Barrier bucket gymnastics and transversely split proton beams:
  Performance at the cern proton and super proton synchrotrons.
\newblock {\em Phys. Rev. Accel. Beams}, 25:050101, May 2022.

\bibitem{PhysRevSTAB.9.104001}
S.~Gilardoni, M.~Giovannozzi, M.~Martini, E.~M\'etral, P.~Scaramuzzi,
  R.~Steerenberg, and A.-S. M\"uller.
\newblock Experimental evidence of adiabatic splitting of charged particle
  beams using stable islands of transverse phase space.
\newblock {\em Phys. Rev. ST Accel. Beams}, 9:104001, 2006.

\bibitem{PhysRevSTAB.12.014001}
A.~Franchi, S.~Gilardoni, and M.~Giovannozzi.
\newblock Progresses in the studies of adiabatic splitting of charged particle
  beams by crossing nonlinear resonances.
\newblock {\em Phys. Rev. ST Accel. Beams}, 12:014001, 2009.

\bibitem{Borburgh:2137954}
J.~Borburgh, S.~Damjanovic, S.~Gilardoni, M.~Giovannozzi, C.~Hernalsteens,
  M.~Hourican, A.~Huschauer, K.~Kahle, G.~Le~Godec, O.~Michels, and
  G.~Sterbini.
\newblock {First implementation of transversely split proton beams in the
  {CERN} Proton Synchrotron for the fixed-target physics programme}.
\newblock {\em EPL}, 113(3):34001. 6 p, 2016.

\bibitem{PhysRevAccelBeams.20.061001}
A.~Huschauer, A.~Blas, J.~Borburgh, S.~Damjanovic, S.~Gilardoni,
  M.~Giovannozzi, M.~Hourican, K.~Kahle, G.~Le~Godec, O.~Michels, G.~Sterbini,
  and C.~Hernalsteens.
\newblock Transverse beam splitting made operational: Key features of the
  multiturn extraction at the {CERN Proton Synchrotron}.
\newblock {\em Phys. Rev. Accel. Beams}, 20:061001, 2017.

\bibitem{PhysRevAccelBeams.20.014001}
S.~Abernethy, A.~Akroh, H.~Bartosik, A.~Blas, T.~Bohl, S.~Cettour-Cave,
  K.~Cornelis, H.~Damerau, S.~Gilardoni, M.~Giovannozzi, C.~Hernalsteens,
  A.~Huschauer, V.~Kain, D.~Manglunki, G.~M\'etral, B.~Mikulec, B.~Salvant,
  J.-L. Sanchez~Alvarez, R.~Steerenberg, G.~Sterbini, and Y.~Wu.
\newblock {Operational performance of the {CERN} injector complex with
  transversely split beams}.
\newblock {\em Phys. Rev. Accel. Beams}, 20:014001, 2017.

\bibitem{PhysRevAccelBeams.22.104002}
A.~Huschauer, H.~Bartosik, S.~Cettour Cave, M.~Coly, D.~Cotte, H.~Damerau,
  G.~P. Di~Giovanni, S.~Gilardoni, M.~Giovannozzi, V.~Kain,
  E.~Koukovini-Platia, B.~Mikulec, G.~Sterbini, and F.~Tecker.
\newblock {Advancing the {CERN} proton synchrotron multiturn extraction towards
  the high-intensity proton beams frontier}.
\newblock {\em Phys. Rev. Accel. Beams}, 22:104002, Oct 2019.

\bibitem{DeLellis2015}
G.~De~Lellis.
\newblock Search for hidden particles (ship): a new experiment proposal.
\newblock {\em Nucl. Part. Phys. Proc.}, 263–264(12):71--76, Oct 2015.

\bibitem{Alekhin_2016}
S.~Alekhin, W.~Altmannshofer, T.~Asaka, B.~Batell, F.~Bezrukov, K.~Bondarenko,
  A.~Boyarsky, K.-Y. Choi, C.~Corral, N.~Craig, and et~al.
\newblock A facility to search for hidden particles at the {CERN} {SPS}: the
  ship physics case.
\newblock {\em Rep. Prog. Phys.}, 79(12):124201, Oct 2016.

\bibitem{Hancock:234723}
S.~Hancock, M.~Martini, M.~Van~Rooij, and C.~Steinbach.
\newblock {Experience with a fast wire scanner for beam profile measurements at
  the CERN PS}.
\newblock Technical report, CERN, Geneva, Apr 1991.

\bibitem{Agoritsas:319333}
V.~Agoritsas, E.~Falk, F.~Hoekemeijer, J.~Olsfors, and C.~Steinbach.
\newblock {The fast wire scanner of the CERN PS}.
\newblock Technical report, CERN, Geneva, Mar 1995.

\bibitem{Steinbach:275924}
C.~Steinbach.
\newblock {Emittance measurements with the CERN PS wire scanner}.
\newblock Technical report, CERN, Geneva, Jan 1995.

\bibitem{ref:pCVD}
{CIVIDEC Instrumentation GmbH}.
\newblock {B2 pCVD Diamond Detector}.

\bibitem{Regenstreif:214352}
E.~Regenstreif.
\newblock {\em The {CERN} Proton Synchrotron}.
\newblock {CERN} Yellow Reports: Monographs. {CERN}, Geneva, 1959.
\newblock French version published as {CERN} 58-06.

\bibitem{Regenstreif:102347}
E.~Regenstreif.
\newblock {\em The {CERN} Proton Synchrotron}.
\newblock {CERN} Yellow Reports: Monographs. {CERN}, Geneva, 1960.
\newblock French version published as {CERN} 59-26.

\bibitem{Regenstreif:278715}
E.~Regenstreif.
\newblock {\em The {CERN} Proton Synchrotron}.
\newblock {CERN} Yellow Reports: Monographs. {CERN}, Geneva, 1962.
\newblock French version published as {CERN} 61-09.

\bibitem{Regenstreif:342915}
E.~Regenstreif.
\newblock {The {CERN} proton synchrotron, 1}.
\newblock {\em Ned. T. Natuurk.}, 25:117--132, 1959.

\bibitem{Regenstreif:342916}
E.~Regenstreif.
\newblock {The {CERN} proton synchrotron, 2}.
\newblock {\em Ned. T. Natuurk.}, 25:149--159, 1959.

\bibitem{Burnet:1359959}
J.-P. Burnet, C.~Carli, M.~Chanel, R.~Garoby, S.~Gilardoni, M.~Giovannozzi,
  S.~Hancock, H.~Haseroth, K.~Hübner, D.~Küchler, J.~Lewis, A.~Lombardi,
  D.~Manglunki, M.~Martini, S.~Maury, E.~Métral, D.~Möhl, G.~Plass,
  L.~Rinolfi, R.~Scrivens, R.~Steerenberg, C.~Steinbach, M.~Vretenar, and
  T.~Zickler.
\newblock {\em Fifty years of the {CERN} Proton Synchrotron: Volume 1}.
\newblock {CERN} Yellow Reports: Monographs. {CERN}, Geneva, 2011.

\bibitem{Steerenberg:EPAC06-MOPCH097}
R.~R. Steerenberg, J.-P. Burnet, M.~Giovannozzi, O.~Michels, E.~Métral, and
  B.~Vandorpe.
\newblock {CERN Proton Synchrotron Working Point Control Using an Improved
  Version of the Pole-face-windings and Figure-of-eight Loop Powering}.
\newblock In {\em Proc. 10th European Particle Accelerator Conf. (EPAC'06)},
  pages 264--266. JACoW Publishing, Jun. 2006.

\bibitem{Blas:IPAC13-WEPME011}
A.~Blas, S.~S. Gilardoni, and G.~Sterbini.
\newblock {Beam Tests and Plans for the CERN PS Transverse Damper System}.
\newblock In {\em Proc. 4th Int. Particle Accelerator Conf. (IPAC'13)}, pages
  2947--2949. JACoW Publishing, May 2013.

\bibitem{Sterbini:HB14-MOPAB10}
G.~Sterbini, A.~Blas, and S.~S. Gilardoni.
\newblock {Beam-based Performance of the CERN PS Transverse Feedback}.
\newblock In {\em Proc. 54th ICFA Advanced Beam Dynamics Workshop on
  High-Intensity and High-Brightness Hadron Beams (HB'14)}, pages 40--44. JACoW
  Publishing, Nov. 2014.

\bibitem{Sterbini:IPAC16-THPMR043}
G.~Sterbini et~al.
\newblock {Performance of Transverse Beam Splitting and Extraction at the CERN
  Proton Synchrotron in the Framework of Multi-turn Extraction}.
\newblock In {\em Proc. 7th Int. Particle Accelerator Conf. (IPAC'16)}, pages
  3492--3495. JACoW Publishing, May 2016.
\newblock https://doi.org/10.18429/JACoW-IPAC2016-THPMR043.

\bibitem{stein:ipac16-mopmr031}
O.~Stein et~al.
\newblock {Investigation of Injection Losses at the Large Hadron Collider with
  Diamond Based Particle Detectors}.
\newblock In {\em Proc. IPAC'16}, pages 310--312. JACoW Publishing, Geneva,
  Switzerland, 2016.

\bibitem{PhysRevAccelBeams.23.044802}
A.~Gorzawski, R.~B. Appleby, M.~Giovannozzi, A.~Mereghetti, D.~Mirarchi,
  S.~Redaelli, B.~Salvachua, G.~Stancari, G.~Valentino, and J.~F. Wagner.
\newblock Probing lhc halo dynamics using collimator loss rates at 6.5 tev.
\newblock {\em Phys. Rev. Accel. Beams}, 23:044802, Apr 2020.

\bibitem{PhysRevAccelBeams.23.124501}
B.~Lindstrom, P.~B\'elanger, A.~Gorzawski, J.~Kral, A.~Lechner, B.~Salvachua,
  R.~Schmidt, A.~Siemko, M.~Vaananen, D.~Valuch, C.~Wiesner, D.~Wollmann, and
  C.~Zamantzas.
\newblock Dynamics of the interaction of dust particles with the lhc beam.
\newblock {\em Phys. Rev. Accel. Beams}, 23:124501, Dec 2020.

\bibitem{FLUKA:2015}
G.~Battistoni, T.~Boehlen, F.~Cerutti, P.~W. Chin, L.~S. Esposito, A.~Fass\'o,
  A.~Ferrari, A.~Lechner, A.~Empl, A.~Mairani, A.~Mereghetti, P.~Garcia~Ortega,
  J.~Ranft, S.~Roesler, P.~Sala, V.~Vlachoudis, and G.~Smirnov.
\newblock {Overview of the FLUKA code}.
\newblock {\em Annals of Nuclear Energy}, 82:10--18, 2015.

\bibitem{FLUKA:2022}
C.~Ahdida, D.~Bozzato, D.~Calzolari, F.~Cerutti, N.~Charitonidis, A.~Cimmino,
  A.~Coronetti, G.~L. D'Alessandro, A.~Donadon~Servelle, L.~S. Esposito,
  R.~Froeschl, R.~Garc\'ia~Al\'ia\, A.~Gerbershagen, S.~Gilardoni,
  D.~Horv{\'a}th, G.~Hugo, A.~Infantino, V.~Kouskoura, A.~Lechner, B.~Lefebvre,
  G.~Lerner, M.~Magistris, A.~Manousos, G.~Moryc, F.~Ogallar~Ruiz, F.~Pozzi,
  D.~Prelipcean, S.~Roesler, R.~Rossi, M.~Sabate~Gilarte, F.~Salvat~Pujol,
  P.~Schoofs, V.~Str{\'a}nsky, C.~Theis, A.~Tsinganis, R.~Versaci,
  V.~Vlachoudis, A.~Waets, and M.~Widorski.
\newblock {New Capabilities of the FLUKA Multi-Purpose Code}.
\newblock {\em Frontiers in Physics}, 9, 2022.

\bibitem{LineBuilder}
A.~Mereghetti, V.~Boccone, F.~Cerutti, R.~Versaci, and V.~Vlachoudis.
\newblock {The FLUKA Linebuilder and Element Database: Tools for Building
  Complex Models of Accelerators Beam Lines}.
\newblock {\em Conf. Proc.}, C1205201:WEPPD071. 4 p, May 2012.

\bibitem{madx}
{MAD - Methodical Accelerator Design}.
\newblock \url{https://mad.web.cern.ch/mad/}.

\bibitem{Schmidt:573082}
F.~Schmidt, E.~Forest, and E.~McIntosh.
\newblock {Introduction to the polymorphic tracking code: Fibre bundles,
  polymorphic Taylor types and "Exact tracking"}.
\newblock Technical report, {CERN}, Geneva, Jul 2002.

\bibitem{Forest:ICAP06-MOM1MP02}
E.~Forest, Y.~Nogiwa, and F.~Schmidt.
\newblock {The FPP and PTC Libraries}.
\newblock In {\em Proc. 9th Int. Computational Accelerator Physics Conf.
  (ICAP'06)}, pages 17--21. JACoW Publishing, Oct. 2006.

\bibitem{CableAttenuation}
E.~Calvo~Giraldo.
\newblock {Private communication}, 2022.

\bibitem{Hernalsteens:IPAC13-WEPEA054}
C.~Hernalsteens et~al.
\newblock {CERN PS Optical Properties Measured with Turn-by-turn Orbit Data}.
\newblock In {\em Proc. 4th Int. Particle Accelerator Conf. (IPAC'13)}, pages
  2627--2629. JACoW Publishing, May 2013.

\bibitem{Skowro:IPAC19-MOPTS102}
P.~K. Skowronski, M.~Giovannozzi, and A.~Huschauer.
\newblock {Linear and Non-Linear Optics Measurements in PS using Turn-by-Turn
  BPM Data}.
\newblock In {\em Proc. 10th Int. Particle Accelerator Conf. (IPAC'19)}, pages
  1114--1117. JACoW Publishing, May 2019.
\newblock https://doi.org/10.18429/JACoW-IPAC2019-MOPTS102.

\bibitem{GiovannozziEPJPlusTransition}
M.~Giovannozzi, L.~Huang, A.~Huschauer, and A.~Franchi.
\newblock A novel non-adiabatic approach to transition crossing in a circular
  hadron accelerator.
\newblock {\em The European Physical Journal Plus}, 136(11):1189, 2021.

\bibitem{Forest:ICAP06-WEPPP04}
E.~Forest, Y.~Nogiwa, and F.~Schmidt.
\newblock {The FPP Documentation}.
\newblock In {\em Proc. 9th Int. Computational Accelerator Physics Conf.
  (ICAP'06)}, pages 191--193. JACoW Publishing, Oct. 2006.

\bibitem{Giovannozzi:PAC03-RPAG012}
M.~Giovannozzi et~al.
\newblock {Optics Studies for the CERN Proton Synchrotron: Linear and Nonlinear
  Modelling Using Beam Based Measurements}.
\newblock In {\em Proc. 20th Particle Accelerator Conf. (PAC'03)}, pages
  2913--2915. JACoW Publishing, May 2003.

\bibitem{PhysRevSTAB.16.051001}
S.~Gilardoni, M.~Giovannozzi, and C.~Hernalsteens.
\newblock {First observations of intensity-dependent effects for transversely
  split beams during multiturn extraction studies at the {CERN} Proton
  Synchrotron}.
\newblock {\em Phys. Rev. ST Accel. Beams}, 16:051001, May 2013.

\bibitem{Giovannozzi:IPAC17-WEPIK088}
M.~Giovannozzi, A.~Huschauer, O.~Michels, A.~Nicoletti, and G.~Sterbini.
\newblock {Analysis of Performance Fluctuations for the CERN Proton Synchrotron
  Multi-Turn Extraction}.
\newblock In {\em Proc. 8th Int. Particle Accelerator Conf. (IPAC'17)}, pages
  3135--3138. JACoW Publishing, May 2017.
\newblock https://doi.org/10.18429/JACoW-IPAC2017-WEPIK088.

\bibitem{edwards_syphers_1993}
D.~A. Edwards and M.~J. Syphers.
\newblock {\em An Introduction to the Physics of High Energy Accelerators}.
\newblock J. Wiley \& Sons, Inc, 1993.

\end{thebibliography}







\end{document}